\documentclass[preprint2]{aastex}
\def\stacksymbols #1#2#3#4{\def\theguybelow{#2}
        \def\verticalposition{\lower#3pt}
        \def\spacingwithinsymbol{\baselineskip0pt\lineskip#4pt}
        \mathrel{\mathpalette\intermediary#1}}
\def\intermediary #1#2{\verticalposition\vbox{\spacingwithinsymbol
        \everycr={}\tabskip0pt
        \halign{$\mathsurround0pt#1\hfil##\hfil$\crcr#2\crcr
                \theguybelow\crcr}}}
\def\lta{\stacksymbols{<}{\sim}{2.5}{.2}}
\def\gta{\stacksymbols{>}{\sim}{3}{.5}}

\begin{document}

\title{INFLUENCE OF COOLED INTERSTELLAR GAS ON 
THE FUNDAMENTAL PLANE FOR ELLIPTICAL GALAXIES}

\author{William G. Mathews$^1$ and Fabrizio Brighenti$^{1,2}$}

\affil{$^1$University of California Observatories/Lick Observatory,
Board of Studies in Astronomy and Astrophysics,
University of California, Santa Cruz, CA 95064\\
mathews@lick.ucsc.edu}

\affil{$^2$Dipartimento di Astronomia,
Universit\`a di Bologna,
via Ranzani 1,
Bologna 40127, Italy\\
brighenti@bo.astro.it}

%\vskip 2.in
%\noindent
%Received:

%\noindent
%PROOFS TO BE SENT TO:

%\noindent
%Lick Observatory

%\noindent
%Santa Cruz, CA 95064

%\noindent
%$^1$UCO/Lick Observatory Bulletin No.

\vskip .2in

\begin{abstract}

We explore here the possibly important influence of cooled 
interstellar gas on the fundamental plane of elliptical galaxies.
Interstellar cooling is described by a parameterized sink 
term in the equation of continuity.
Parameters that give the best fits to the X-ray observations 
of elliptical galaxy NGC 4472 are used as a template 
for other galaxies.
These spatially-dependent cooling parameters are then extended 
homologously to elliptical galaxies of lower mass for which
X-ray observations are currently unavailable 
or impossible to observe because of stronger relative X-ray emission 
from binary stars.
Most of the interstellar 
gas cools within an effective radius where it 
can contribute an additional 10 - 30 percent 
to the mass of the old stellar population.
The absence of observed cooled gas and simple theoretical 
arguments support the hypothesis 
that the cooled gas is forming into 
low mass stars, with implications that depend on the IMF of 
this young stellar population.
If the IMF includes only stars of very low mass, $\ll M_{\odot}$ 
as is commonly 
believed, the cooled mass is optically dark. 
For this IMF the mass to light ratios 
determined from stellar velocities
systematically overestimate 
that of the old stellar population.
Furthermore, the total mass and spatial distribution
of the optically dark stellar population 
does not scale homologously with galactic luminosity
or radius 
so the total stellar mass to light ratio is expected to
vary with galactic radius.
These variations in stellar population and non-homology 
can introduce pronounced 
deviations from the fundamental plane.
We investigate cooled gas 
perturbations to the mass to light ratio 
for several idealized homologous
elliptical galaxies and show that they may 
be incompatible 
with the observed thinness of the fundamental plane.
However, if the IMF of the stellar population 
produced from cooled interstellar gas is optically luminous, 
its influence on observed stellar mass to light ratios 
and the fundamental plane would be lessened. 
If the IMF of a young, continuously forming 
and optically luminous stellar population 
were sufficiently well-behaved, it could 
account for about ten percent of the optical light from
large elliptical galaxies within $r_e$, possibly
having important implications for understanding the 
fundamental plane.

\end{abstract}


\clearpage

\section{INTRODUCTION}

The mass of hot interstellar gas that cools over a Hubble time 
in a massive elliptical galaxy is typically 5 - 7 percent of  
the total baryonic mass of these galaxies.
Several observations indicate that this gas cools in 
a significant volume of the inner galaxy:
(1) the total cooled mass significantly 
exceeds the mass of central massive 
black holes (Magorrian et al. 1998),
(2) although much of the interstellar gas within $r_e$ is 
ejected from old galactic stars which typically have 
a systematic galactic rotation,
the inner X-ray images of rotating elliptical galaxies are nearly  
circular, implying that the gas cools before it 
flows very far 
toward the rotation axis (Hanlan \& Bregman 2000; 
Brighenti \& Mathews 2000b), 
(3) direct evidence of diffuse 
cooled gas at $T \sim 10^4$ K is 
invariably observed in large elliptical galaxies 
in $r \lta 0.3r_e$ (e.g. Macchetto et al. 1996), and 
(4) evolutionary models for the hot interstellar gas 
are in better agreement with X-ray observations when 
cooling mass loss is included; most of this 
gas cools within an effective radius 
(Brighenti \& Mathews 2000a).

In view of 
the very small masses of cold HI or H$_2$ gas that 
have been observed in elliptical galaxies,  
far less than the mass of hot gas that has cooled, 
it is usually assumed that the cooled interstellar gas 
is continuously forming into a population of very low mass 
(brown dwarf) stars (Fabian, Nulsen \& Canizares 1982).
Such stars add mass but no light.
We show here that if the cooled gas is optically dark,
the fundamental plane for elliptical galaxies 
is likely to be perturbed beyond observed limits.
Alternatively,
if the mass to light ratio of the forming stars is 
not infinite but is comparable to that of the old stars, 
then some of the cosmic scatter or tilt 
of the fundamental plane from the virial plane could 
be due to the presence of this young stellar population.
Support for a continuously forming, optically luminous 
stellar population in elliptical galaxies 
is provided by the nearly solar mass of neutral cores 
in interstellar clouds at which gravitational 
collapse first occurs, $\sim 2$ $M_{\odot}$ 
(Mathews \& Brighenti 1999a), and by the apparent young 
stellar ages of many elliptical galaxies 
based on the H$\beta$ photometric 
index (Worthey 1994; Worthey, Trager \& Faber 1996;
Trager 1997; Mathews \& Brighenti 1999b).

In the following discussion, however, we adopt the usual 
hypothesis that newly forming stars 
in elliptical galaxies are optically dark. 
In this case radial variations in the total stellar
mass to light ratio must occur.
Furthermore, the total mass of cooled gas varies 
with galactic luminosity,
resulting in rather profound excursions from the
fundamental plane,
evidently in violation of the remarkable thinness 
and regularity of this plane.

\section{THE FUNDAMENTAL PLANE}

If elliptical galaxies were perfectly homologous stellar systems 
with identical stellar populations, then the 
virial theorem $\sigma_o^2 = \kappa_v (M/r_e) 
= \kappa_v (L_V/r_e)(M/L_V)$ would imply simple 
correlations among observable parameters.
Here $\sigma_o$ is the ``central'' velocity dispersion,
$M$ is the stellar (or galactic) mass, $r_e$ is the effective  
(half-light) radius and $\kappa_v$ is a constant of 
proportionality.
The three observables -- $\sigma_o^2$, $L_V$, and $r_e$ -- define 
a homologous plane (HP) in this 
(logarithmic) three-dimensional parameter space.
Instead, elliptical galaxies satisfy a different, 
slightly steeper fundamental plane (FP) 
relation $\sigma_o^2 \propto (L_V/r_e)[r_e^{0.22}\sigma_o^{0.49}]$,
implying that $\kappa_v (M/L_V) \propto r_e^{0.22} \sigma_o^{0.49}
\propto {M}^{0.24} r_e^{-0.02} \propto {L_V}^{0.32} r_e^{-0.03}$
(Dressler et al. 1987; Djorgovski \& Davis 1987).
In view of the importance of elliptical galaxies for $D_n - \sigma_o$ 
distance determinations, 
considerable effort has gone into understanding 
the physical origins of the ``tilt'' of the FP relative to 
the HP.

Interpretations of this tilt are constrained by the regularity
of the FP-HP deviation across the entire region observed
and the remarkably small observational 
scatter ($\sim 12$\%) about the FP relation (Renzini \& Ciotti 1993).
Generally, the FP-HP tilt 
has been interpreted as 
some combination of two limiting effects: 
either 
(1) stellar populations or dark matter 
contributions vary systematically across the FP so that 
$M/L_V \propto {L_V}^{0.32}$ or 
(2) homology fails so that $\kappa_v \propto {L_V}^{0.32}$. 
This relation depends mildly on photometric band 
(e.g. Scodeggio et al. 1998; Pahre \& Djorgovski 1997).

Several types of stellar population variation  
(age, metallicity or IMF) 
have been considered along the FP 
(Renzini \& Ciotti 1993;
Pahre, Djorgovski \& Carvalho 1995; Pahre \& Djorgovski 1997). 
These attempts to interpret the FP-HP tilt 
have not been entirely successful since  
the FP is nearly invariant to observed 
wavelength over the $B - H$ photometric bands.  
The small systematic wavelength trend of the FP slope is 
consistent with the color-magnitude 
relation for giant elliptical galaxies (Scodeggio et al 1998;
Pahre, de Carvalho \& Djorgovski 1998). 
Moreover, if the stellar population changes across the FP, 
differential stellar evolution can in some cases 
introduce a change 
in the FP slope at large redshifts, but no such effect 
has yet been observed.
It is generally asserted that the full FP tilt 
cannot be understood solely by metallicity variations 
or changing stellar populations along the FP since 
this would require a high degree of fine tuning 
in the IMF.
If $M/L$ is assumed to be constant with galactic 
radius, the mass to light ratios for elliptical galaxies 
found from dynamic models vary in a manner similar 
to the FP deviation, $M/L \propto L^{0.35 \pm 0.05}$
(van der Marel 1991).

A wide variety of non-homologous structural and/or kinematic 
models have been proposed to explain all or part of the FP-HP tilt
(e.g., Capelato et al. 1995;
Djorgovski 1995; Hjorth \& Madsen 1995; 
Ciotti, Lanzoni \& Renzini 1996).
In these types of models the
stellar $M/L_V$ does not vary so the FP tilt should be 
strictly independent of wavelength apart from 
mutually non-homologous stellar and dark matter distributions. 
Graham \& Colless (1997) fit the stellar mass and light distribution 
with Sersic $r^{1/n}$ laws and note that $n$ increases 
with galactic mass (by $\sim 2-4$) 
in a non-homologous fashion across the FP. 
If these galaxies are mapped onto the FP with 
an $r^{1/4}$ law in the usual way, the resulting FP, 
$r_e \propto \sigma_o^{1.33 \pm 0.10} \Sigma_e^{-0.79 \pm 0.11}$,
tilts even further from the HP, 
$r_e \propto \sigma_o^{2} \Sigma_e^{-1}$, 
where $\Sigma_e \propto L/r_e^2$ is the surface brightness.
However, a dynamical non-homology must accompany a  
structural non-homology.  
Solving the Jeans equation for elliptical galaxies 
modeled with $r^{1/n}$ profiles
(Ciotti 1991; Ciotti \& Lanzoni 1997), 
Graham \& Colless (1997) and 
Graham (1998) determine infinite-aperture
$\sigma_{tot,n}$ which when fit to $r^{1/n}$ galaxies result in
$r_{e,n} \propto \sigma_{tot,n}^{1.72 \pm 0.24} \Sigma_e^{-0.74 \pm
0.09}$, which is somewhat closer to the HP.
Busarello et al. (1997) consider a different dynamic non-homology 
based on partial rotational support within $r_e$. 
Using a modified $\tilde{\sigma}$ 
which incorporates both rotation and 
random velocities, they find that more than half of the FP-HP tilt is 
accounted for, 
$r_e \propto \tilde{\sigma}^{1.53 \pm 0.2} \Sigma_e^{-0.92 \pm 0.1}$.
Busarello et al. claim that 
the remaining discrepancy can be understood 
with spatial non-homology and stellar population differences 
(e.g. Prugniel \& Simien 1997). 

Another approach is the possibility that 
dark (halo) matter makes a variable contribution along the FP 
(Ciotti, Lanzoni, \& Renzini 1996), resulting in the FP-HP tilt. 
However, 
these authors stress again the tightness of observational 
scatter about the FP so that any systematic trend in 
the dark matter content must be rather finely tuned. 
This high degree of coordination 
may seem unlikely if elliptical galaxies are produced by a 
variety of random mergers among stochastically 
different galaxies of lesser mass.  
The amount and radial distribution of dark halo matter 
in the centers of elliptical galaxies is complicated 
by the baryonic winds in first generation galaxies and the 
dissipational baryonic concentration expected as giant 
elliptical galaxies form by mergers; these 
counteracting effects respectively 
contract and expand the dark matter distribution 
relative to the baryonic stellar population.

There is at present 
no clear winner among the various explanations for the FP-HP tilt. 
Instead, a patchwork of small 
contributions -- stellar population variations, 
rotational support and non-homology -- 
are required to bring FP into agreement with HP
(Ciotti, Lanzoni, \& Renzini 1996: 
Prugniel \& Simien 1997;
Pahre, de Carvalho \& Djorgovski 1998).
However, there is no compelling reason why these 
quite different and presumably independent 
contributions should 
combine in a coordinated fashion 
to produce the full FP-HP tilt observed across 
the entire FP. 
Moreover, a multiplicity of coordinated 
explanations for the tilt, 
none entirely dominant, 
makes the tightness of the FP even more remarkable. 

\section{COOLING DROPOUT MASS AND LOW MASS STARS}

In this paper we introduce a new complication to the 
FP-HP controversy: a variation in 
the dynamical $M/L_B$ with galactic mass 
due to the deposition of cooled interstellar gas in $r \lta r_e$. 
The importance of cooled gas in the galactic mass budget 
follows from the universality of hot interstellar gas in 
giant elliptical galaxies and the direct relationship of the 
cooled mass to the loss of energy implied by 
the observed X-ray emission $L_x$.
For example, 
interstellar gas in the large Virgo elliptical 
galaxy NGC 4472  
with temperature $T \approx 1.3 \times 10^7$ K 
emits $L_x(0.5 - 4.5 {\rm keV}) = 4.5 \times 10^{41}$ 
ergs s$^{-1}$ in the {\it Einstein} band. 
The bolometric X-ray luminosity,
$L_{x,bol} \approx 1.6 L_x(0.5 - 4.5 {\rm keV})$, 
is the product of the 
mass cooling rate ${\dot M}$ and the specific enthalpy of the 
gas, therefore 
\begin{equation}
{\dot M} = \left({2 \mu m_p \over 5 k T}\right) L_{x,bol}
\approx 2.5 M_{\odot}~{\rm yr}^{-1}.
\end{equation}
Since $L_x$ was somewhat larger in the past, 
over a Hubble time $\sim 3 - 4 \times 10^{10}$ $M_{\odot}$ 
of hot gas will have cooled. 
Although this amounts to only about $\sim 5 - 10$ percent of the 
total stellar mass in NGC 4472, 
$M_{*t} = 4.73 \times 10^{11}$ $M_{\odot}$, 
most of the cooled gas is likely to be deposited 
near the galactic center $r \lta r_e$ where the 
hot gas density and X-ray emissivity are greatest 
and where its contribution to the dynamical $M/L_B$ 
determined from stellar velocity dispersions is 
most apparent. 
(In evaluating $M_{*t}$ we assume 
$M/L_B = 6$, as explained below.) 

Although there is little doubt that interstellar 
cooling has occurred, 
two main questions must be addressed: (1) What is the final physical 
state of the cold gas? and (2) What is the radial distribution 
of the cooled mass in the galaxy? 
For many years it has been assumed that low mass stars 
are the most likely final state for cooled gas 
(e.g. Fabian, Nulsen \& Canizares 1982).
In a recent paper (Mathews \& Brighenti 1999a) we 
describe how the hot interstellar gas 
cools (initially) to $T \sim 10^4$ K where it 
can be heated and ionized 
by stellar UV photons 
and made visible in optical emission lines. 
This first stage of cooling occurs at many ($\sim 10^6$) 
cooling sites distributed over the inner galaxy. 
However, 
when the radius of partially cooled gas at $10^4$ K 
exceeds the local Stromgren radius 
at a cooling site, a cold neutral 
atomic or molecular core forms  
with $T_c \sim 15$ K and $n_c \sim 10^5$ cm$^{-3}$ 
in pressure equilibrium with the ambient 
hot interstellar gas. 
As the mass of the neutral core $M_c$ increases further, 
it becomes Jeans unstable when $M_c \sim 1 - 2$ $M_{\odot}$ 
so only stars of this or lower mass 
can form from the available gas 
(Ferland, Fabian \& Johnstone 1994; 
Mathews \& Brighenti 1999a). 
Cooling gas is processed within 
HII cooling site clouds with star-forming cold cores 
in time scales ($\sim 10^5$ years) that 
are short compared to their hydrodynamic disruption times 
in the cooling flow atmosphere 
(Mathews \& Brighenti 1999a).
The high efficiency of low mass star formation nicely 
accounts for the absence of observable cold HI or molecular gas 
in large elliptical galaxies, e.g. for 
NGC 4472 $M_{HI} + M_{H2} < 10^7$ $M_{\odot}$.
We conclude that 
two stellar populations coexist in giant elliptical galaxies: 
an old luminous stellar component with an IMF similar 
to Salpeter, and a second population of low mass stars
continuously forming from the interstellar medium 
and having a bottom heavy IMF. 
Although the detailed character of this second IMF is 
unknown, for the purposes of the following discussion 
we adopt the usual assumption 
that it is restricted to very low masses, i.e. 
its mass to light ratio is essentially infinite.

The second question -- concerning 
the radial distribution of 
cooled gas mass -- can be answered in an approximate 
way by comparing hydrodynamic models with the 
radial properties of interstellar gas observed in X-rays.
The production of low mass stars at many distributed 
sites can only occur if the interstellar gas is inhomogeneous. 
Radiative cooling is locally enhanced in regions of 
low entropy (low $T$, high density $n$). 
As gas in the galactic cooling 
flow slowly proceeds inward, 
regions of progressively higher entropy cool at 
smaller galactic radii; regions having low entropy 
cool further from the galactic centers.
Distributed H$\alpha$ + [NII] line emission 
observed in the central 
regions of giant elliptical galaxies illuminate the 
cooling sites where this process is occurring.
Interstellar entropy variations can be 
created by a variety of complex events: 
stellar disturbances (e.g. stellar winds or Type Ia supernovae), 
interactions with merging dwarf galaxies 
or differential Type II supernova heating in the early 
universe during the earliest phases of star formation. 
Because of our poor understanding of these interstellar 
entropy fluctuations, 
it is not possible to determine from first principles 
where most of the hot gas cools and deposits its mass into 
low mass stars, although the observed optical line emission 
is a visible tracer of this process. 
The best currently available 
means of setting limits on the radial distribution 
of cooling gas in elliptical galaxies 
is to compare the observed radial 
distribution of interstellar X-ray emission 
with the computed X-ray emission from 
gas-dynamic models for the cooling flow gas in which cooling 
dropout is assumed to occur according to various 
radial prescriptions. 
We have recently computed a series of models 
for the large elliptical galaxy NGC 4472 with a variety 
of assumed mass dropout functions
(Brighenti \& Mathews 2000a).
In the following section we describe the cooling dropout 
properties of models that best fit the X-ray observations,
then we discuss the implications of this dropout mass 
for the fundamental plane.

\section{DESCRIPTION OF HYDRODYNAMICAL MODELS}

In view of the uncertainty concerning the cooling 
dropout, in our gas dynamical models 
of galactic cooling flows we parameterize the radial 
mass dropout of cooling gas
by introducing an adjustable sink term in the 
equation of continuity,
\begin{equation}
{ \partial \rho \over \partial t}
+ {1 \over r^2} { \partial \over \partial r}
\left( r^2 \rho u \right) = \alpha \rho_*
-q(r) {\rho \over t_{do}}.
\end{equation}
Here $\rho(r,t)$ and $u(r,t)$ are the gas density and velocity 
in the cooling flow.
The sink term for cooling flow dropout is 
characterized by a dimensionless parameter $q(r)$ 
and $t_{do} = 5 m_p k T / 2 \mu \rho \Lambda$
is the time for gas at temperature $T(r,t)$ 
to cool locally by radiative
losses at constant pressure (see, e.g. Sarazin \& Ashe 1989).
The source term $\alpha(t) \rho_*(r)$ represents the rate that 
interstellar gas is supplied by stellar mass loss from 
the old stellar population having density 
$\rho_*(r)$.
Since $\alpha(t) \propto (M/L_B)^{-1}$ for a wide 
range of power law IMFs,
the total stellar mass ejected $\int \alpha(t) M_* dt$ 
is nearly identical for 
elliptical galaxies with the same $L_B$ but different IMFs
(Brighenti \& Mathews 2000a).
Following the approach in 
Brighenti \& Mathews (2000a), 
we consider galactic cooling flow models 
for NGC 4472 using a parameterized dropout coefficient 
\begin{equation}
q(r) = q_o \exp(-r/r_{do})^m
\end{equation}
where $q_o$, $r_{do}$ and $m$
are adjustable parameters.
For all sets of dropout parameters considered 
we found that the total dropout mass 
in $r \lta r_e/3$ is a substantial fraction (0.10 - 0.35) of the 
mass of luminous stars in this region.  
The radial distribution of the dropout mass, assumed to 
remain at the cooling site, does not 
follow the space density  
corresponding to the de Vaucouleurs light profile. 

If low mass stars formed from cooling dropout 
contribute mass but no appreciable light, 
the hypothesis that we want to test,
it follows that 
the intrinsic ``stellar'' mass to light ratio of the 
luminous stellar population in NGC 4472 must be 
somewhat lower than values
determined from the observed motion of
luminous stars: 
$M/L_B \approx 9.2$ (van der Marel 1991) 
or $M/L_B \approx 10.3$ (Faber et al. 1997).
The most satisfactory 
mass dropout coefficients $q(r)$ generate hydrodynamic  
solutions for the interstellar gas 
at the current time ($t_n = 13$ Gyrs) 
that reproduce best the observed X-ray image of NGC 4472. 
Using this approach, 
we find that the dynamical mass to light ratio in 
successful hydrodynamic models 
requires a somewhat lower  
intrinsic mass to light ratio for the luminous stars  
$M/L_B \approx 6$. 
For the low mass 
dropout stellar population we assume 
$(M/L_B)_{do} = \infty$. 

Ideally, it would be desirable to determine if the 
same (or appropriately scaled) mass dropout coefficient $q(r)$
used in gas dynamical models for elliptical galaxies of varying $L_B$,
spread across the fundamental plane, 
can fit the observed X-ray surface brightness distributions 
$\Sigma_x(R)$ in these same galaxies.
Unfortunately, at the present pre-{\it Chandra} time sufficiently  
accurate $\Sigma_x(R)$ are known only for several bright 
elliptical galaxies in Virgo 
(e.g., Trinchieri, Fabbiano \& Canizares 1986), 
all having similar $L_B$. 
Approximate $\Sigma_x(R)$ distributions are available only for 
about a dozen elliptical galaxies altogether (Mathews \& Brighenti 1998a).
In addition, emission from low mass binary stars 
begins to dominate the X-ray luminosity in elliptical 
galaxies several times less luminous than 
NGC 4472, masking the X-ray emission from the cooling flows.
Finally, even though $L_x \propto L_B^2$ 
approximately holds 
among brighter elliptical galaxies, 
there is an enormous spread in values of $L_x$ for given $L_B$.
We believe that most of 
this spread is due to environmental interactions: 
(i) tidal exchange 
of halo hot gas (and dark matter) in the group environments 
in which these elliptical galaxies formed (Mathews \& Brighenti 1998b)
or (ii) ram pressure stripping. 

With these difficulties in mind,
we chose instead to explore the relative influence of 
optically dark interstellar dropout not directly on the FP but on 
a homologous galactic plane (HP) that is scaled from NGC 4472
for which good X-ray observations are available.
We consider a 
sequence of three massive elliptical galaxies, NGC 4472 and two similar 
model galaxies, 4472/4 and 4472/16,  
having optical luminosities $L_B$  
that are respectively 4 and 16 times less than NGC 4472. 
To explore only the structural non-homology caused by cooling dropout, 
we also assume that the old stellar population is invariant 
along our HP, i.e. its $M/L_B$ is constant.
For perfect homology we also require that the
mass and scale radius of the dark halo distribution scale 
with the baryonic mass.
Finally, we solve for the evolution of hot gas in each galaxy  
along the HP using identical (or scaled) 
parameters ($q_o$, $r_{do}$ and $m$) that 
give good agreement with the X-ray observations for NGC 4472,
$T(r,t_n)$ and $n(r,t_n)$; this procedure is described in 
more detail in Brighenti \& Mathews (2000a).
By this means we can determine the differential, non-homologous 
distribution of optically dark dropout mass and assess the 
deviation of these three idealized 
galaxies from the HP relation.
We expect that these deviations from the HP are 
similar to deviations of galaxies of similar $L_B$ 
from the local FP.
If our procedure were perfectly correct -- i.e., elliptical 
galaxies are exactly homologous and contain identical 
old stellar 
populations -- we would find that the non-homology introduced 
by cooling dropout 
would explain the ``tilt'' of the observed FP from the HP.
Unfortunately, this goal has not yet been achieved.

For perfect homology and simplicity we assume that 4472 and its 
two clones of lower mass are spherical with 
de Vaucouleurs stellar profiles and scaled NFW  
dark halos (Navarro, Frenk, \& White 1996). 
The effective radius of NGC 4472 is $r_e = 8.57$ kpc (for distance  
$d = 17$ Mpc) and its total (old) stellar mass is 
$M_{*,t} = 4.73 \times 10^{11}$ $M_{\odot}$, 
based on $M/L_B = 6$. 
The NFW halo in NGC 4472 has a virial mass of 
$M_{h,vir} = 4 \times 10^{13}$ $M_{\odot}$.
These parameters give a good fit to the X-ray 
mass profile of NGC 4472 assuming hydrostatic equilibrium 
although the NFW profile is found to be 
somewhat too centrally peaked (Brighenti \& Mathews 2000a).
For the smaller comparison galaxies
(4472/4 and 4472/16) we choose identical
stellar parameters but scale the size of the stellar
distribution downward by $4^{-0.793} = 0.333$
or $16^{-0.793} = 0.111$
where the exponent is found from the
observed $M_V$-$r_e$ relation for massive elliptical galaxies,
$r_e \propto L_V^{0.793} \propto
L_B^{0.793}$, (Faber et al. 1997).
The virial mass of the NFW halos for the smaller galaxies
are reduced by 4 or 16 and the scale radius for these
halos varies as $r_s \propto M_{h,vir}^{0.457}$,
as required by our chosen cosmology for which the NFW 
concentration varies as $c \propto M_{h,vir}^{-0.123}$ 
(Navarro, Frenk, \& White 1996).

As explained in more detail elsewhere 
(Brighenti \& Mathews 1999a),
our gas dynamical models begin with an overdensity in a simple 
Einstein-de Sitter cosmology ($\Omega = 1$, 
$\Omega_b = 0.05$, and $H_o = 50$  km s$^{-1}$ Mpc$^{-1}$)
toward which baryonic and dark matter are gravitationally 
attracted.
Galactic stars are assumed to form at 
time $t_{*s} = 1$ Gyr and, soon afterwards,
Type II supernova energy 
based on a Salpeter IMF (slope: $x = 1.35$, mass limits:
$m_{\ell} = 0.08$ and $m_u = 100$ $M_{\odot}$) 
heats the ambient gas within the  
accretion shock with 80 percent efficiency 
(see Brighenti \& Mathews 1999a for details).
At $t_* = 2$ Gyrs the de Vaucouleurs stellar profile 
and potential is constructed, simulating the results 
of mergers.
For this approximate exploration
we ignore the possibility that elliptical galaxies
of lower luminosity may have formed slightly earlier.
In some models we allow the secondary inflow of 
cosmic gas to continue throughout the calculation. 
For comparison, in other models 
we ignore the secondary infall entirely, crudely simulating a 
tidal truncation and removal 
of halo material in a relatively dense group of newly formed 
galaxies, 
a process more likely in elliptical galaxies of moderate or low mass.

\section{RESULTS}

In Table 1 we list the main results of a variety of 
hydrodynamic cooling flow models for 4472, 4472/4 and 4472/16 
with various assumed dropout parameters and 
assumed amounts of accumulated cosmic gas. 
Of particular interest are the mass to light ratios 
evaluated at $r_e/3$ and at a fixed distance of 480 pc 
from the galactic centers.  
Recall that $M/L_B = 6$ is assumed 
for the luminous stars in all three 
galaxies and that the observed value for 
NGC 4472 is $\sim 9.2$ (van der Marel 1991) 
or $\sim 10.3$ (Faber et al. 1997).
Models in which the (stellar plus dropout) 
mass to light ratio for NGC 4472 
falls in the range $M/L_B \approx 8.5 - 11$ 
can be regarded as 
reasonably successful. 
However, the $M/L_B(r_e/3)$ entries in Table 1 are 
enhanced by about 15 percent due to a small amount of 
NFW dark halo material in this region. 
Since our fits to the X-ray surface brightness of NGC 4472 
are improved if the dark halo is less centrally peaked than NFW 
(also see Kravtsov et al. 1998; Burkert \& Silk 1997;
Navarro \& Steinmetz 2000), 
we believe that its contribution
to $M/L_B$ in this region should be removed.
Therefore, for comparison with 
$M/L_B(r_e/3)$ values in Table 1 which 
contain this NFW excess, 
the range of acceptable models for NGC 4472   
should be revised upward to 
$M/L_B(r_e/3) \approx 9.8 - 12.6$
(i.e. $\log [M/L_B(r_e/3)] = 0.99 - 1.10$). 
Values of $M/L_B$ at 480 pc may be more relevant 
for comparison with $M/L_B$ determinations that 
are often based on 
stellar velocity dispersions measured within 
a fixed aperture of a few 
arcseconds about the galactic centers.

In cooling flows without distributed dropout ($q_o = 0$) 
the flow proceeds 
entirely to the origin before cooling and the mass to light 
ratios at 480 pc greatly exceed observed values 
for 4472 -- such models can certainly be rejected. 
Evolutionary gas dynamic models with uniform $q_o = 1$ at all radii 
and with dropout parameters 
($q_o,~r_{do},~m$) = (4, 2 kpc, 1) 
both fit the observed distribution of interstellar gas temperature 
and density in NGC 4472 reasonably well at the 
present time ($t_n = 13$ Gyrs) as illustrated in Figure 1.
While the agreement in Figure 1 
is acceptable, it is by no means perfect.
The apparent gas density (enhanced by dropout emission) exceeds 
observed densities in the central regions 
($r \lta 3$ kpc) of NGC 4472, 
suggesting that some additional non-thermal support is 
present; NGC 4472 has a weak extended radio source 
in this region with 
an appreciable equipartition pressure 
(Ekers \& Kolanyi 1978). 
At large galactic radii ($r \gta 10$ kpc) 
our models in Figure 1 are slightly lower than 
the azimuthally-averaged values observed in NGC 4472; 
this density mismatch can be corrected with small 
changes in the galaxy formation time $t_*$ 
or the efficiency of SNII energy release 
(Brighenti \& Mathews 1999a).
However, our interest here is not to achieve 
absolute perfection in fitting 
the observations of NGC 4472, but to make a differential 
comparison of 
4472 with its less massive, homologous clones; 4472/4 
and 4472/16.
The additional model with stronger but uniform 
dropout $q = q_o = 4$, also included 
in Table 1, results in apparent 
gas temperatures at $t_n = 13$ Gys 
that are considerably lower than those observed
(Brighenti \& Mathews 2000a); this is due 
to enhanced X-ray emission from cooling regions throughout 
the cooling flow.

Several interesting details are apparent in Table 1. 
Since the dropout mass is deposited non-homologously, 
the dynamic mass to light ratio varies 
with galactic radius, 
i.e. $M(r_e/3)/L_B \neq M(480~{\rm pc})/L_B$.
Even when the dropout parameters are scaled homologously  
from 4472 
[($q_o,~r_{do},~m$) = (4, 2 kpc, 1)]
to 4472/4
[($q_o,~r_{do},~m$) = (4, 0.667 kpc, 1)],
the resulting $M/L_B$ are {\it not} equal, 
i.e. dropout does not scale homologously. 
Some models in Table 1 were performed 
with no additional secondary infall gas following 
the time $t_*$ of galaxy formation 
(NO in column 5).
This procedure is justified by the strong 
correlation of normalized X-ray sizes $r_{ex}/r_e$
with $L_x/L_B$ noted by Mathews \& Brighenti (1998).
We explained this correlation 
with models of tidally truncated galactic dark halos 
and (mostly secondary infallen) halo gas in smaller galaxies 
and the transfer of this material to halos of 
more massive, centrally located elliptical galaxies, 
a process which is dynamically expected in galaxy 
groups where elliptical galaxies are formed.
The removal of secondary infallen 
halo gas is therefore more relevant to 
4472/4 and 4472/16 than for 4472 (which may have acquired 
additional halo gas from neighboring galaxies); 
note that this late arriving cosmic  
gas also has a larger influence on $M/L_B$ 
for the smaller galaxies.

The non-homologous dropout behavior listed 
in Table 1 is illustrated in 
Figure 2 in which the 4472 $\rightarrow$ 4472/4 
$\rightarrow$ 4472/16 variation 
of $M/L_B(r_e/3)$ and $M/L_B(480~{\rm pc})$ 
are plotted for each model. 
The local fundamental plane, 
$M/L_B \propto {L_B}^{\beta}$, shown with dashed lines, 
has a slope $\beta = 0.24$ 
that is intermediate between values 
for the B-band FP 
taken from Jorgensen, Franx, \& Kjaergaard (1996) and 
Scodeggio et al. (1998).
The dashed lines are normalized to 
$M/L_B = 9.2$ (from van der Marel 1991; long dashed line) and 
$M/L_B = 10.3$ (from Faber et al. 1997; short dashed line)
respectively. 
The asterisks show the constant stellar $M/L_B = 6$ assumed 
for luminous stars in all three galaxy models. 
Clearly, none of the model pairs 
exhibit a variation of $M/L_B(r_e/3)$ 
as steep as the FP relation, 
but the dropout contribution to the FP slope is 
nevertheless consequential at this radius.
However, the variation of 
$M/L_B$ at $r = 480$ pc for 
the $q_o = 1$ models follows the fundamental 
plane quite accurately over the range of $L_B$ between 
4472 and 4472/4. 
Note that the pair of variable dropout models
with homologously scaled parameters 
[$(q_o,~r_{do},~m) = (4, 2~{\rm kpc}, 1)$ for 4472 and 
$(q_o,~r_{do},~m) = (4, 0.667~{\rm kpc}, 1)$ for 4472/4] 
produce $M/L_B(480)$ that 
vary in a sense opposite to the FP trend, 
but only when all cosmic gas from secondary infall is included. 
Our mass to light values for 4472 may be lower limits since 
we have not considered models in which unusually massive galaxies 
receive more than their cosmological allotment of halo material 
by tidal acquisitions from less massive, 
unmerged group galaxies.

The constant $q = q_o = 1$ models can explain 
much of the FP tilt in 4472 $\rightarrow$ 4472/4 
with or without secondary infall.
However, when extended to 4472/16, values of 
$M/L_B(480)$ clearly exceed the fundamental plane 
for all models. 
While the dropout contribution to $M/L_B(480)$ 
in the $q = 1$ model (without cosmic inflow)
decreases with $L_B$ over the range of 16 
shown in Figure 2 (i.e. vertical distance between 
open squares and asterisks), 
the dropout mass does not disappear 
entirely for the 4472/16 model. 
In fact the
assumed $M/L_B = 6$ for the luminous 
stars alone (asterisks in Fig. 2) also lies above the FP
for 4472/16.
Within the framework of our optically dark dropout
assumption, this indicates that an independent variation 
of $M/L_B$ for the old stars would be required 
in addition to that provided by the dropout.
Results shown in the upper panel of Figure 2 suggest 
that the FP could be fit over the 
full 4472 $\rightarrow$ 
4472/16 range if we had chosen a somewhat lower value 
of $M/L_B$ for the luminous stars, 
but the X-ray data for NGC 4472 become difficult 
to fit if $M/L_B \lta 4$.

However, dynamically significant 
galactic rotation is expected for elliptical 
galaxies having B-band luminosities less than 
that of 4472/4, $2 \times 10^{10}$
$L_{B,\odot}$, which lies near the transition between massive,
boxy elliptical galaxies with cores and low mass, rotationally
supported, disky elliptical galaxies with power law
profiles (Faber et al. 1997).
Rotation can reduce the mass dropout in smaller galaxies 
since the cooled gas will be deposited in a relatively 
larger galactic volume, 
reducing $M/L$ values for 4472/16 in Figure 2.
Although the excess interstellar density in our models 
within 1 - 3 kpc (Figure 1) can be understood in terms 
of $\sim 100~\mu$G magnetic fields, it could also 
indicate that we have somewhat 
overestimated the gas density in 
this region and also the mass of cooled gas. 

Perhaps the best resolution of the large $M/L$ in our 
models of 4472/16 would be to allow some of 
the cooled gas to form a luminous stellar 
population. 
Using this assumption, we have shown that 
the anomalously low ages for bright elliptical galaxies 
found from observations of the H$\beta$ photometric 
index (e.g., Trager 1997) can be quite naturally produced 
by a young population with a Salpeter IMF extending 
up to $\sim 2$ $M_{\odot}$ but no further 
(Mathews \& Brighenti 1999a).
In models having luminous dropout stars, 
about 10 percent of the optical light 
is produced by the young stars.
The stellar disk component characteristic of rotating, 
low luminosity elliptical galaxies 
(Rix, Carollo, \& Freeman 1999), 
can be formed from younger cooling dropout stars
(Brighenti \& Mathews 1997b), 
and there is some evidence that these galaxies 
also have enhanced H$\beta$ indices
(de Jong \& Davies 1997).

In our models we determine the detailed radial 
variation of the mass to light ratio within $r_e$;
this variation depends most sensitively on our choice of $q(r)$.
In general, optical observations of stellar velocity dispersion  
profiles cannot determine the detailed variation of 
$M/L$ with radius within $r_e$.
Only with considerable effort have observers been able to 
detect the presence of dark matter at $r \gta r_e$
(e.g., Carollo et al. 1995; Rix et al. 1997), although
this dark matter is immediately 
apparent from X-ray observations (e.g. Brighenti \& Mathews 1997a).
The insensitivity of optical 
observations to the detailed radial variation of $M/L$
is probably due to the generic stellar 
orbital bias toward radial anisotropy in $r \lta r_e$; 
this results in 
$M/L$ determinations that more nearly represent 
orbital averages rather than local values.

For completeness, 
in the final column of Table 1 we list the ROSAT X-ray luminosity 
of each of our models. 
In general our results are slightly flatter than the 
observed trend for luminous elliptical galaxies, 
$L_x \propto L_B^p$ with $p \approx 2$, 
in which the X-ray emission is interstellar 
(e.g. Eskridge, Fabbiano \& Kim 1995).
However, we have shown elsewhere 
(Mathews \& Brighenti 1998) that $L_x/L_B$ correlates with 
the size of the X-ray image and that this correlation 
can be explained by the tidal transfer of halo material 
in galaxy groups from subordinate elliptical galaxies to the 
dominant central elliptical in the group.
Galactic rotation can also influence
$L_x/L_B$.
Because of these environmental influences, 
the values of $L_x$ in Table 1 should not 
be regarded as a strong test of a particular model.

\section{FINAL REMARKS AND CONCLUSIONS}

The full extent of the fundamental plane 
for elliptical galaxies spans a range of $L_B$ 
about $\sim 3$ times greater than the factor  
of 16 shown in Figure 2
(Burstein et al. 1997). 
Maintaining the FP-HP tilt 
over this larger range would be difficult if cooling dropout 
is an important contributor of dark mass. 
Nevertheless, the presence of appreciable dropout mass 
seems unavoidable. 

The total mass of cooled interstellar gas present in 
elliptical galaxies 
could be reduced by supernova-driven 
galactic winds or by the formation of stars that have since evolved.
For our models we assume a Type Ia rate
SNu$(t) =~$ SNu$(t_n)$$(t_n/t)^p$ supernovae per 100 years
per $10^{10}$ $L_B$ with $p = 1$ and
SNu$(t_n) = 0.03$ (Brighenti \& Mathews
2000a).
Winds produced by more numerous Type Ia supernovae at 
early times (i.e. $p \gta 1.3$) 
cannot be invoked to expel some of the dropout 
material from galaxies like 4472/4 or 4472/16 
since these same higher Type Ia 
rates in more massive galaxies would 
greatly overproduce iron in the current interstellar gas, 
in violation of X-ray observations.
Our value SNu$(t_n) = 0.03$ was chosen to be 
consistent with the low interstellar 
iron abundances for elliptical 
galaxies reported in most of the recent literature,
$z_{Fe,ism} \lta 0.4$ solar (e.g., Tanaka et al. 1994).
If the abundances are more nearly solar (Buote 1999), 
as is likely, 
then a proportionally larger SNu$(t_n)$ is indicated; this 
can result in galactic winds in the 4472/16 model 
at late times $t \sim t_n$.
Therefore, the total mass of interstellar gas that cools 
can be reduced by 
SNII-driven galactic outflow at early times 
(when most of the dropout occurs) and (less efficiently) by 
SNIa-driven outflow at late times.
However, the large $L_x$ observed in virtually all 
bright elliptical galaxies clearly rules out 
ongoing galactic winds in these galaxies.
The current amount of dropout mass could also be 
reduced if the IMF of the forming stars was 
more luminous in the past (i.e. more massive stars). 
While none of these means of reducing the dropout mass is 
securely understood, some significant mass of cooled 
gas must be present.
The large mass of cooled interstellar gas 
expected within $r_e$ together with the 
alarming number of possible ways of enhancing or 
moderating the amount of cooled gas 
makes the coherency and tightness
of the FP quite baffling.

Our main objective in presenting these 
results is to emphasize the importance 
of cooling dropout mass to the FP-HP controversy.
The results that we have shown here are based on the 
commonly held assumption that only 
optically dark stars of very low mass can form in cooling flows.
However, optically dark dropout by itself appears to be 
an unpromising explanation 
for the full FP tilt, although with some well chosen parameters 
that might be possible. 
What we see in Figure 2 is an unwelcome and quite 
unaesthetic disturbance of the $M/L$ relation, 
one that involves the uncertain physics of star formation  
and the origin of entropy fluctuations in the hot gas.

Nevertheless, the mass contribution from cooled interstellar gas 
cannot be dismissed by those concerned with the deviation 
of the fundamental plane from a truly homologous plane.
Like other effects proposed to explain the FP-HP 
discrepancy, if the cooling dropout mass is a dominant 
contributor, the mass dropout must 
vary in a smooth, regular fashion across 
the FP, proportional to some power of $L_B$.
Such regularity could conceivably 
result from the similar profiles of 
interstellar gas density and temperature observed in the central 
regions of elliptical galaxies ($r \lta r_e$), provided the entropy 
fluctuations are also similar. 
Our apparent success in explaining the the anomalously young ages 
of elliptical galaxies indicated by the H$\beta$ index 
from an optically luminous dropout stellar population 
(Mathews \& Brighenti 1999b) strongly suggests that 
a better match to the FP-HP tilt could be achieved if the 
optically dark assumption used here is relaxed.
In this case the IMF of the continuously forming stars 
is likely to extend to higher maximum masses in elliptical 
galaxies of lower 
$L_B$ because of the lower interstellar pressure in low luminosity 
elliptical galaxies.

\acknowledgments

Studies of the evolution of hot gas in elliptical galaxies 
at UC Santa Cruz is supported by
grants from NASA and the NSF 
for which we are very grateful. In addition FB is supported
in part by Grant MURST-Cofin 98.

%\clearpage

%\centerline{\bf APPENDIX}
%\appendix

%\end{document}

\clearpage

\makeatletter
\def\jnl@aj{AJ}
\ifx\revtex@jnl\jnl@aj\let\tablebreak=\nl\fi
\makeatother

\footnotesize

\begin{deluxetable}{lcccccrrr}
%%%\tabletypesize{\scriptsize}
\normalsize
\tablenum{1}
\tablewidth{40pc}
%\tablewidth{0pc}
\tablecaption{COOLING DROPOUT PARAMETERS AND MASS TO LIGHT RATIOS}
\tablehead{
\colhead{GALAXY} &
\colhead{$L_B$} &
\colhead{$q_o$} &
\colhead{$r_{do}$\tablenotemark{a}} &
\colhead{$m$\tablenotemark{a}} &
\colhead{COSMIC\tablenotemark{b}} &
\colhead{${M\over L_B}({r_e \over 3})$} &
\colhead{~~${M\over L_B}(480\; {\rm pc})$} &
\colhead{~~~~$L_x$\tablenotemark{c}} \cr
\colhead{} &
\colhead{$(10^{10} L_{B,\odot})$} &
\colhead{} &
\colhead{(kpc)} &
\colhead{} &
\colhead{GAS?} &
\colhead{} &
\colhead{} & 
\colhead{~~~~$(10^{40}$ ergs s$^{-1})$} \cr
}
\startdata
4472/16: & &  &     &     &       &         &        & \cr
 &0.493  &  0  &  \nodata  &  \nodata  &  NO   &  9.32  &  8.57 
& 0.10 \cr
 &0.493 &  1  &  $\infty$  &  \nodata  &  NO   &  7.29  & 7.13 
& 0.10 \cr
 &0.493 &  4  &  0.317  &  1  &  NO   &  8.25  &  8.17 
& 0.25 \cr
4472/4: & &  &     &     &       &         &        & \cr
 &1.97  &  0  &  \nodata  &  \nodata  &  YES  &  11.76  &  19.27 
& 9.35 \cr
 &1.97  &  0  &  \nodata  &  \nodata  &  NO   &  8.97  &  12.37 
& 1.56 \cr
 &1.97  &  1  &  $\infty$  &  \nodata  &  YES  &  7.95  & 8.41 
& 8.00 \cr
 &1.97  &  1  &  $\infty$  &  \nodata  &  NO   &  7.40  & 7.65 
& 1.52 \cr
 &1.97  &  4  &  0.667  &  1  &  YES  &  11.05  &  10.91 
& 14.57 \cr
 &1.97  &  4  &  0.667  &  1  &  NO   &  8.65  &  8.54 
& 2.57 \cr
 &1.97  &  4  &  2  &  1  &  YES  &  8.16  & 7.27  
& 10.78 \cr
 &1.97  &  4  &  $\infty$  &  \nodata  &  YES  &  7.03  & 6.79  
& 6.00 \cr
4472: & &  &     &     &       &         &        & \cr
 &7.89  &  0  &  \nodata  &  \nodata  &  YES  &  10.47  & 32.29  
& 20.08 \cr
 &7.89  &  0  &  \nodata  &  \nodata  &  NO   &  9.97  & 28.53  
& 8.59 \cr
 &7.89  &  1  &  $\infty$  &  \nodata  &  YES  &  8.83  & 11.40 
& 31.56 \cr
 &7.89  &  1  &  $\infty$  &  \nodata  & NO    &  8.64  & 10.96 
& 12.56 \cr
 &7.89  &  4  &  2  &  1  &  YES  &  9.74  & 7.99  
& 28.11 \cr
 &7.89  &  4  &  2  &  1  &  NO   &  9.37  & 7.88  
& 13.73 \cr
 &7.89  &  4  & 0.8 &  2  &  YES  & 10.47  & 13.08 
& 36.44 \cr
 &7.89  &  4  &  $\infty$   &  \nodata  &  YES  & 7.88  & 7.20 
& 39.82 \cr
\enddata
\tablenotetext{a}{When $q(r) = q_o$ is constant, 
the exponential factor in equation (3) is set to unity, as if 
$r_{do} = \infty$ and $m$ is  undefined.}
\tablenotetext{b}{YES (NO) indicates that continued secondary flow of
cosmic gas into the galactic halo is (not) included.}
\tablenotetext{c}{ROSAT band X-ray luminosity (including dropout 
emission) within 150 kpc for NGC 4472, 50 kpc for 4472/4 
and 16.6 kpc for 4472/16}
\end{deluxetable}

\normalsize

\clearpage

\vskip.1in
\figcaption[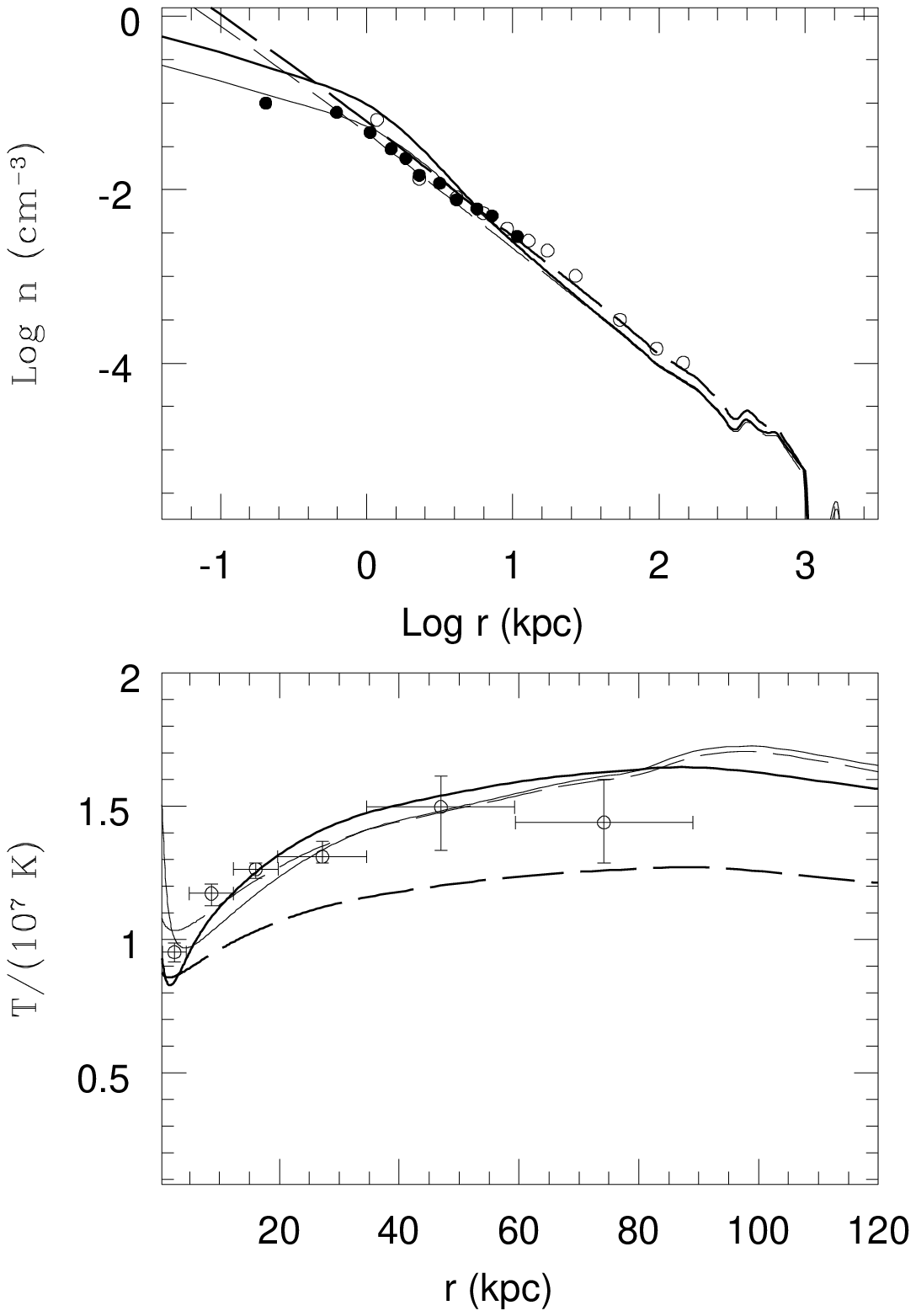]{
{\it Top panel:} Comparison of radial interstellar density variation 
for two hydrodynamic models at time $t_n = 13$ Gyrs 
with observed density in NGC 4472.
Filled circles are {\it Einstein} data
(Trinchieri, Fabbiano, \& Canizares 1986), open circles 
are from ROSAT (Irwin \& Sarazin 1996).
The light solid and dashed lines refer to the background 
(uncooled) gas density in the $q_o = 1$ and  
$(q_o,~r_{do},~m) = (4, 2~{\rm kpc}, 1)$ models respectively.
The heavy lines are the apparent space variation of the 
gas density including emission from gas that is cooling 
locally. 
{\it Bottom panel:} Comparison of the interstellar temperature 
variation with radius computed in the same two models 
with ROSAT temperature data (Irwin \& Sarazin 1996). 
The light solid and dashed lines refer to the computed 
temperature variation in the uncooled background flow 
with physical radius. 
The heavy solid and dashed lines show the apparent variation 
of temperature projected on the sky and corrected for 
emission from mass dropout; these curves should be compared 
with the observations. 
\label{fig1}}

\vskip.1in
\figcaption[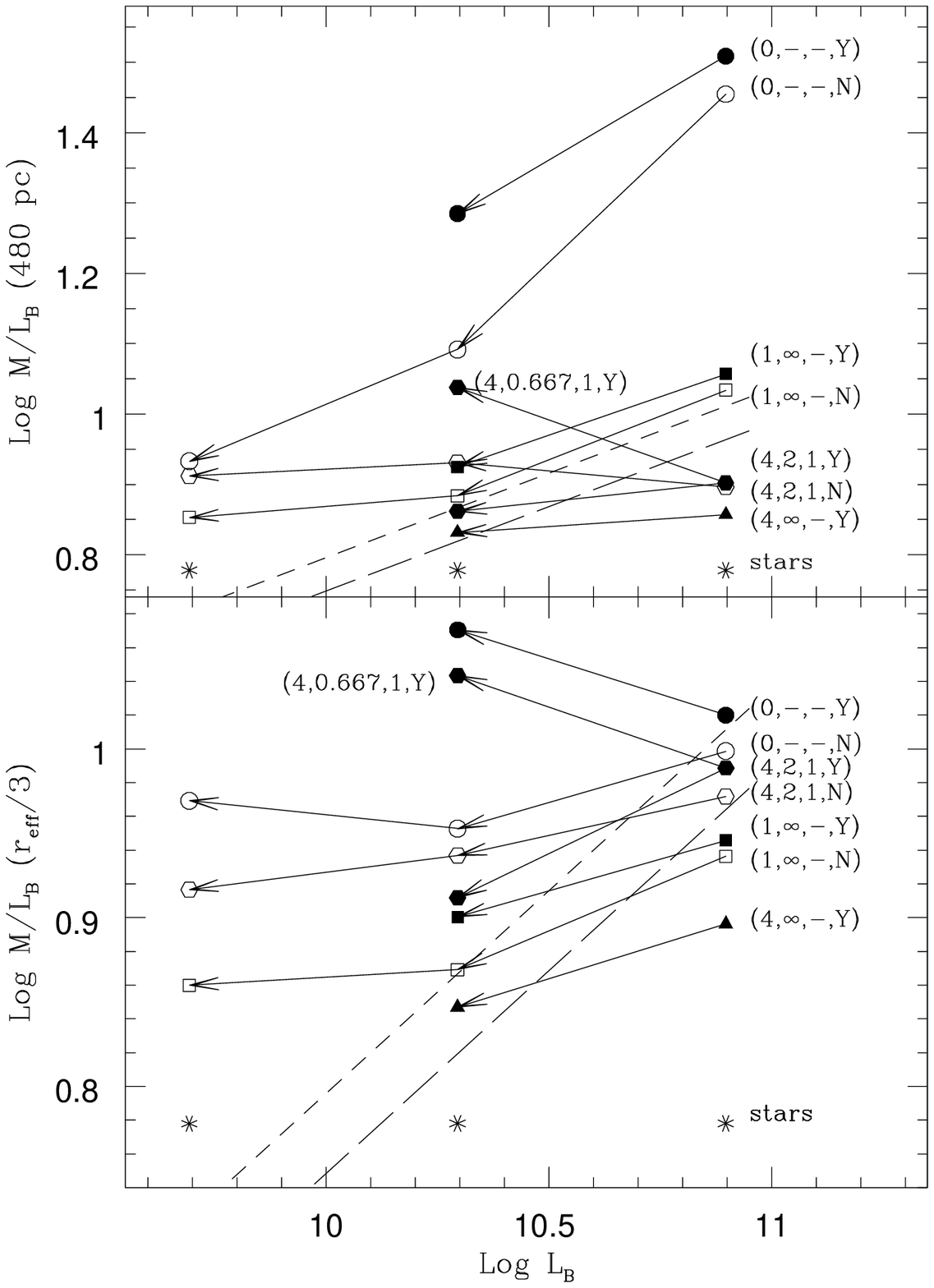]{
Variation of the mass to light ratio $M/L_B$ 
with luminosity $L_B$ (both in solar units)
at $t_n = 13$ Gyrs.
$M/L_B$ includes both  
luminous stars and interstellar dropout for NGC 4472 (on the 
right) and two homologously scaled galaxies 
having $1/4$ and $1/16$  
the mass of NGC 4472 (center and left respectively). 
{\it Top panel:} $M/L_B$ evaluated 
within fixed radius $r = 480$ pc for all galaxies;
{\it Bottom panel:} $M/L_B$ within $r = r_e/3$ for all galaxies. 
Asterisks show $M/L_B = 6$ assumed for the luminous stars.
Dashed lines show the local 
the fundamental plane, $M/L_B \propto {L_B}^{0.24}$,
normalized at NGC 4472 
with $M/L_B = 9.2$ ({\it long dashed line}) 
or $M/L_B = 10.3$ ({\it short dashed line}).
Circles refer to no-dropout models ($q_o = 0$),
squares to uniform $q_o = 1$ models,
triangles to uniform $q_o = 4$ models and 
hexagons to variable dropout models with 
$(q_o,~r_{do},~m) = (4, 2~{\rm kpc}, 1)$ for 4472 and   
$(q_o,~r_{do},~m) = (4, 0.667~{\rm kpc}, 1)$ 
for 4472/4 (upper filled hexagon) or 
$(q_o,~r_{do},~m) = (4, 2~{\rm kpc}, 1)$ for 4472/4 
(lower filled hexagon).
Gas from secondary cosmic infall is included in solutions 
with filled symbols and this gas is excluded in solutions 
with open symbols. 
Where possible the symbols are labeled with $(q_o,~r_{do},~m,~Y/N)$; 
$Y/N$ refers to the presence or lack of secondary infall.
\label{fig2}}

\end{document}